\documentclass[runningheads]{llncs}
\usepackage[T1]{fontenc}
\usepackage{amsmath}
\usepackage{amssymb}
\usepackage{fancyvrb}       % for fancy verbatime options like size
\usepackage[most]{tcolorbox}
\usepackage{inconsolata}
\usepackage{xcolor}
\usepackage{tikz}
\usepackage{enumitem}
\usepackage{hyperref}
\usepackage{graphicx}
\usepackage{everysel}
\usepackage{float}
\usepackage{caption}
\usepackage{gensymb}
\usepackage{tikz}
\usetikzlibrary{arrows.meta}

\spnewtheorem*{appendixthm}{Theorem}{\bfseries}{\itshape}
\begin{document}

% Author macros::begin %%%%%%%%%%%%%%%%%%%%%%%%%%%%%%%%%%%%%%%%%%%%%%%%
\title{Formally Verified Trades in Financial Markets} %optional, in case that the title is too long; the running title should fit into the top page column
\author{Suneel Sarswat \and Abhishek Kr Singh}

\institute{Tata Institute of Fundamental Research, Mumbai\\ \email{\{suneel.sarswat, abhishek.uor\}@gmail.com}}

\newcommand{\tw}{\texttt}
\newcommand{\rw}{\rightarrow}

\maketitle

\begin{abstract}
We introduce a formal framework for analyzing trades in financial markets.  These days, all big exchanges use computer algorithms to match buy and sell requests and these algorithms must abide by certain regulatory guidelines. For example, market regulators enforce that a matching produced by exchanges should be \emph{fair}, \emph{uniform} and \emph{individual rational}. To verify these properties of trades, we first formally define these notions in a theorem prover and then develop many important results about matching demand and supply. Finally, we use this framework to verify properties of two important classes of double sided auction mechanisms. All the definitions and results presented in this paper are completely formalized in the Coq proof assistant without adding any additional axioms to it. 
\end{abstract}

\section{Introduction}
\label{section1}

In this paper, we introduce a formal framework for analyzing trades in financial markets. Trading is a principal component of all modern economies. Over the past few centuries, more and more complex instruments are being introduced for trade in the financial markets. All big stock exchanges use computer algorithms to match buy requests (demand) with sell requests (supply) of traders. Computer algorithms are also used by  traders to place orders in the markets (known as \emph{algorithmic trading}). With the arrival of computer assisted trading, the volume and liquidity in the markets have increased drastically, and as a result, the markets have become more complex.

Software programs that enable the whole trading process are extremely complex and have to meet high efficiency criteria. Furthermore, to increase the confidence of traders in the markets, the market regulators set stringent safety and fairness guidelines for these software. Traditionally, to meet such criteria, software development has extensively relied on testing the programs on large data sets. Although testing is helpful in identifying bugs, it cannot guarantee the absence of bugs. Even small bugs in the trading software can have a catastrophic effect on the overall economy. An adversary might exploit a bug to his benefit and to the disadvantage of other genuine traders. These events are certainly undesirable in a healthy economy. 

Recently, there have been various instances \cite{nse,nyse2,nyse1} of violation of the trading rules by the stock exchanges. For example, in \cite{nyse1}, a regulator noted: "NYSE Arca failed to execute a certain type of limit order under specified market conditions despite having a rule in effect that stated that NYSE Arca would execute such orders"\footnote{The New York Stock Exchange and the Archipelago Exchange merged together to form NYSE Arca, which is an exchange where both stocks and options are traded.}. This is an instance of a program not meeting its specification. Here the program is a matching algorithm used by the exchange and the regulatory guidelines are the broad specifications for the program. Note that, in most of the cases, the guidelines stated by the regulators are not a complete specification of the program. Moreover, there is no formal guarantee that these  guidelines are consistent. These are some serious issues potentially compromising the safety and integrity of the markets. 

Recent advances in formal methods in computer science can be put to good use in ensuring safe and fair financial markets. During the last few decades, formal method tools have been increasingly successful in proving the correctness of large software and hardware systems \cite{seqcir,blast,compcert,sel4}. While model checking tools have been used for the verification of hardware, the use of interactive theorem provers have been quite successful in the verification of large software. A formal verification of financial algorithms using these tools can be helpful in the rigorous analysis of market behavior at large. The matching algorithms used by the exchanges (venues) are at the core of the broad spectrum of algorithms used in financial markets. Hence, a formal framework for verifying matching algorithms can also be useful in verifying other algorithms used in financial markets. This need has also been recognized by Passmore and Ignatovich \cite{PassmoreI17}. They state \begin{quote} Indeed, if venues are not safe, fair and correct, e.g., if one can exploit flaws in the venue matching logic to jump the queue and have their orders unfairly prioritized over others, then ``all bets are off'' as one ascends the stack to more complex algorithms. 
\end{quote} 

In this work, we make significant progress in addressing this need, including completely formalizing the matching algorithm used in the pre-markets. Before we describe our full contribution, we first briefly describe trading at an exchange. 

\subsection{An Overview of Trading at an Exchange}

An exchange is an organized financial market. There are various types of exchanges: stock exchange, commodity exchange, foreign exchange etc. An exchange facilitates trading between buyers and sellers for the products which are registered at the exchange. A potential trader, a buyer or a seller, places orders in the markets for a certain product. These orders are matched by the stock exchange to execute trades.  Most stock exchanges hold trading in two main sessions: pre-market (or call auction session) and continuous  market (or regular trading session) (See \cite{harris} for details on the market microstructure). 

The pre-market session reduces uncertainty and volatility in the market by discovering an opening price of the product. During the pre-market session, an exchange collects all the buy requests (bids) and sell requests (asks) for a fixed duration of time. At the end of this duration the exchange matches these buy and sell requests at a single price using a matching algorithm. In the continuous market session, the incoming buyers and sellers are continuously matched to each other. An incoming bid (ask), if matchable, is immediately matched to the existing asks (bids). Otherwise, if the bid (ask) is not matchable, it is placed in a priority queue prioritized first by price and then by time. A trader can place orders of multiple quantity of each product to trade during both the sessions. In the continuous market session, unless otherwise specified, an order of multiple units of a product can be partially executed, that too potentially at different trade prices. In the pre-market session, an order of multiple units can always be partially executed and all trades occur at a single price, namely the opening price. In this work, we will be concerned primarily with the pre-market session where orders can always be partially executed, which is also the case for most orders in the continuous market session. Hence, for simplicity of analysis, it suffices to assume that each order is of a single unit of a single product; a multiple quantity order can always be treated as a bunch of orders each with a single quantity and the analysis for a single product will apply for all the products individually. As a result, note that a single trader who places an order of multiple units is seen as multiple traders ordering a single unit each. In both sessions of trades multiple buyers and sellers are matched simultaneously. A mechanism used to match multiple buyers and sellers is known as a double sided auction \cite{friedman}.

In double sided auctions, an auctioneer (e.g. exchanges) collects buy and sell requests over a period of time. Each potential trader places the orders with a limit price: below which a seller will not sell and above which a buyer will not buy. The exchange at the end of this time period matches these orders based on their limit prices. This entire process is completed using a double sided auction matching algorithm. Designing algorithms for double sided auctions is well studied topic \cite{mcafee1992,WurmanWW98,NiuP13}. A major emphasis of many of these studies have been to either maximize the number of matches or  maximize the profit of the auctioneer. In the auction theory literature, the profit of an auctioneer is defined as the difference between the limit prices of matched bid-ask pair. However, most exchanges today earn their profit by charging transaction costs to the traders. Therefore, maximizing the number of matches increases the profit of the exchange as well as the liquidity in the markets.  There are other important properties, like fairness, uniformity and individual rationality, besides the number of matches which are considered while evaluating the effectiveness of a matching algorithm. However, it is known that no single algorithm can possess all of these properties \cite{WurmanWW98,mcafee1992}. 

\subsection{Our Contribution}
Our main goal through this work is to show effectiveness of formal methods in addressing real needs in financial markets and hopefully, along with subsequent works, this will lead to fully-verified real trading systems. In this work, we formally define various notions from auction theory relevant for the analysis of trades in financial markets. We define notions like bids, asks and matching in the Coq proof assistant. The dependent types of Coq turn out to be very useful in giving concise
representation to these notions, which also reflects their natural definitions. After preparing the basic framework, we define important properties of matching in a double sided auction: fairness, uniformity and individual rationality. These properties reflect various
regulatory guidelines for trading. Furthermore, we formally prove some results on the existence of various combinations of these properties. For example, a maximum matching always exists which is also fair. These results can also be interpreted as consistency proofs for
various subsets of regulatory guidelines. We prove all these results in the constructive setting of the Coq proof assistant without adding any additional axioms to it. These proofs are completed using computable functions which computes the actual instances (certificate). We also use computable functions to represent various predicates on lists. Finally, we use this setting to verify properties of two important classes of matching algorithms: uniform price and maximum matching algorithms.

We briefly describe the main results formalized in this work. To follow the discussion below, recall that each bid and each ask is of a single quantity, and hence the problem of pairing bids and asks can be seen as a matching problem between all bids and all asks with additional price constraints.

\noindent \textbf{Upper bound on matching size:} After formalizing the various notions, we first show that these definitions are also useful in formalizing various theorems on double sided auctions by formalizing a combinatorial result (Theorem~\ref{lem:boundM}) which gives a tight upper bound on the total number of possible trades (cardinality of a maximum matching). For a given price, the demand (supply) is the total number of buyers (sellers) willing to trade at that price. Theorem~\ref{lem:boundM} states that for any price $p$, the total number of trades is at most the sum of the demand and supply at price $p$. In order to prove Theorem~\ref{lem:boundM}, we first formalize Lemmas 1-3.

\noindent \textbf{Properties of matchings:} We next formalize theorems relating to three important properties of matchings: fairness, uniformity and individual rationality. Before explaining the theorems, we first explain these terms. 

A matching is {\it unfair} if there exists two buyers who had different bids and the lower bid buyer gets matched but not the higher bid one. Similarly, it could be unfair if a more competitive seller is left out. If a matching is not unfair, then it is {\it fair}. 

A matching is {\it uniform} if all trades happen at the same price and is {\it individually rational} if for each matched bid-ask pair the trade price is between the bid and ask limit prices. In the context of formal markets, the trade price is always between the limit prices of the matched bid-ask pair. Note that, during the pre-market session, a single price is discovered, and thus the exchange is required to produce a uniform matching for this session of trading. 

Theorem~\ref{thrm:IR} states that there exists an algorithm that can convert any matching into individual rational. This can be achieved by assigning the trade prices as the middle values between the limit prices of matched bid-ask pairs. 

Theorem~\ref{thm:fairMatching} states that given a matching there exists a fair matching of the same cardinality. We use two functions \emph{Make\_FOB} and \emph{Make\_FOA} which successively makes the matching fair on the bids and then the asks, thus resulting in a fair matching. The proof of Theorem~\ref{thm:fairMatching}, which is based on induction, uses Lemmas 4-9 and is quite technically subtle, as induction fails when we try to use it directly (see the discussion below Lemma~\ref{lem:fobcorrect}), and we need to first prove intermediate Lemmas \ref{lem:fobcorrect} and \ref{lem:projectsub} before we can use induction. In addition, we exhibit (see Fig.~\ref{fig:mmum}) individual rational matchings to show that they cannot be both uniform and maximum simultaneously.

\noindent \textbf{Matching Algorithms:} Finally, we formalize two important matching algorithms: \emph{produce\_MM} and \emph{produce\_UM}.

Theorem~\ref{thm:maxMatching} shows that produce\_MM always outputs a maximum matching. Composing \emph{Make\_FOB}, \emph{Make\_FOA} (from Theorem~\ref{thm:fairMatching}) and \emph{produce\_MM} (Theorem \ref{thm:maxMatching}), we can show that there exists an algorithm that outputs a maximum matching which is also fair (Theorem~\ref{thm:existfairMM}).

The \emph{produce\_UM} algorithm is implemented by the exchanges for opening price discovery, and Theorem~\ref{thm:uniformMM} states that \emph{produce\_UM} outputs a maximum-cardinality matching amongst all uniform matchings. We can compose \break \emph{Make\_FOA}, \emph{Make\_FOB} (Theorem~\ref{thm:existfairMM}) and \emph{produce\_UM} (Theorem~\ref{thm:uniformMM}) to get an algorithm that produces a maximum matching amongst all uniform matchings that is also fair. Instead, we directly prove that the matching produced by \emph{produce\_UM} is also fair by first proving Lemmas \ref{thm:uniformfairbid} and \ref{thm:uniformfairask}. This completely formalizes the matching algorithm used by the exchanges during the pre-market session of trading.

Finally we observe that while our work is useful for continuous markets, it does not completely formalize trades during the continuous market session. This requires further work as the lists continuously get updated during this session of trading and the order types are also more involved. See the discussion in Conclusion and Future Works (Section~\ref{sec:conclusion}). 

\subsection{Related Work}\label{sec:related}
There is no prior work known to us which formalizes double-sided auction mechanism used by the exchanges. Passmore and Ignatovich in \cite{PassmoreI17} highlight the significance, opportunities and challenges involved in formalizing financial markets. Their work describes in detail the whole spectrum of financial algorithms that need to be verified for ensuring safe and fair markets. Matching algorithms used by the exchanges are at the core of this whole spectrum. Another important work in formalization of trading using model checking tools is done by Iliano \emph{et al.} \cite{clf}. They use concurrent linear logic (CLF) to outline two important properties of a trading system: the market is never in a locked-or-crossed state, and the trading always take place at best bid or best ask limit price. They also highlight the limitation of CLF in stating and proving properties of trading systems. 

On the other hand, there are quite a few works formalizing various concepts from auction theory \cite{toolbox,lange,frank}. Most of these works focus on the Vickrey auction mechanism. In a Vickrey auction, there is a single seller with different items and multiple buyers with valuations for each subset of item. Each buyer places bids for every combination of the items. At the end of the bidding, the aim of the seller is to maximize the total value of the items by suitably assigning the items to the buyers. Financial derivatives and other type of contracts are also formalized in \cite{contract1,contract2}. 

\subsection{Organization of the paper}
In Section~\ref{sec:modeling}, we formally define the essential components of trading at an exchange. In particular, we define some important properties of matchings and prove Theorems 1-3. In Section~\ref{sec:optimization}, we present a maximum matching algorithm (\emph{produce\_MM}) which produces a maximum matching which is fair. We also present an equilibrium price matching algorithm (\emph{produce\_UM}) which is used for price discovery in financial markets. We also specify and prove some correctness properties for these algorithms (Theorems 4-7). We summarize the work in Section~\ref{sec:conclusion} with an overview of future works. The Coq code for this work is available at \cite{auctiongithub}, which can be compiled on the latest version of Coq (8.10.1). Some proof explanations have been moved to the appendix to meet the space constraint.

\section{Modeling Trades at Exchanges}\label{sec:modeling}
An auction is a  competitive event, where goods and services are sold to the most competitive participants. The priority among participating traders is determined by various attributes of the bids and asks (e.g. price, time etc). This priority can be finally represented by ordering them in a list.

\subsection{Bid, Ask and Limit Price}
In  any double sided auction multiple buyers and sellers place their orders to buy or sell a unit of an underlying product. The auctioneer matches these buy-sell requests  based on their \emph{limit prices}. While the limit price for a buy order (i.e. \emph{bid}) is the price above which the buyer does not want to buy the item, the limit price of a sell order (i.e. \emph{ask}) is the price below which the seller does not want to sell the item. If a trader wishes to buy or sell multiple units, he can create multiple bids or asks with different \emph{ids}. We can express bids as well asks using  records containing two fields. 

\begin{Verbatim}[fontsize=\footnotesize]
  Record Bid: Type:= Mk_bid { bp:> nat; idb: nat }.
  Record Ask: Type:= Mk_ask { sp:> nat; ida: nat }.
\end{Verbatim}

For a bid $b$, $(bp \; b)$  is the limit price and $(idb \; b)$ is its unique identifier. Similarly for an ask $a$, $(sp \; a)$ is the limit price and $(ida \; a)$ is the unique identifier of $a$. Note that the limit prices are natural numbers when expressed in the monetary unit of the lowest denomination (like cents in USA). Also note the use of coercion \tw{ :>} in the first field of $Bid$ which declares $bp$ as an implicit function that is applied to any term of type $Bid$ appearing in a context requiring a natural number. Hence from now on we can simply use $b$ instead of ($bp \;b$) for the limit price of $b$. Similarly, we use $a$ for the limit price of an ask $a$.

Since equality for both the fields of $Bid$ as well as $Ask$ is decidable (i.e. \tw{nat: eqType}), the equality on $Bid$ as well as $Ask$ can also be proved to be decidable. This is achieved by declaring two canonical instances \tw{bid\_eqType} and \tw{ask\_eqType} which connect $Bid$ and $Ask$ to the \tw{eqType}.  

\subsection{Matching Demand and Supply}
All the buy and sell requests can be assumed to be present in list $B$ and list $A$, respectively. At the time of auction, the auctioneer matches bids in $B$ to asks in $A$. We say a bid-ask pair $(b, a)$ is \emph{matchable} if $b \ge a$ (i.e. $bp \; b \ge sp \; a$).  Furthermore, the auctioneer assigns a trade price to each matched bid-ask pair which results in  a matching $M$. We define a matching as a list whose entries are of type \tw{fill\_type}.

\begin{Verbatim}[fontsize=\footnotesize]
  Record fill_type: Type:=  Mk_fill {bid_of: Bid; ask_of: Ask; tp: nat} 
\end{Verbatim}

\begin{figure}[h]
\begin{center}
\resizebox {.85\textwidth} {!} {
\begin{tikzpicture}[font=\Large]
%Loop for Lists, top elements, fill colors. 
%Loop for letter A' and M
\foreach \x/\t/\y/\m in {2.2/black/37/.4,3.9/red/69/.4,5.2/blue/82/-.4,5.3/black/83/.4,6.1/green/91/-.4,8.2/orange/112/.4,9/black/120/.4,9.5/magenta/125/.4}
{
 \draw (\x, 2.5) node {$]$};
 \draw [color=\t] (\x, 0) node {$]$};

 \draw (\x, 2.5+\m) node {{\scriptsize \y}};
 \draw [color =\t] (\x, \m) node {{\scriptsize \y}};
}

\foreach \x/\t/\y/\m in {3.3/red/53/.4,4.9/blue/79/.4,5.5/black/85/-.4,6/green/90/.4,6.4/black/94/.4,6.8/orange/98/-.4,8.3/magenta/113/-.4,9.1/black/121/-.4}
{
 \draw (\x, 1.25) node {$[$};
 \draw [color=\t] (\x, 0) node {$[$};

 \draw (\x, 1.25+\m) node {{\scriptsize \y}};
 \draw [color =\t] (\x, \m) node {{\scriptsize \y}};
}

\draw (2,0) -- (10,0);
 
\draw (2,1.25) -- (10,1.25);

\draw (2,2.5) -- (10,2.5);

\draw (1.75,0) node {{\scriptsize $M$}};
 
\draw (1.75,1.25) node {{\scriptsize $A$}};

\draw (1.75,2.5) node {{\scriptsize $B$}};

\end{tikzpicture}
}
\end{center}
\caption{ Bids in $B$ and asks in $A$ are represented using close and open brackets respectively, and a matched bid-ask pair in $M$ is assigned the same colors. }
\label{fig:matching}
\end{figure}
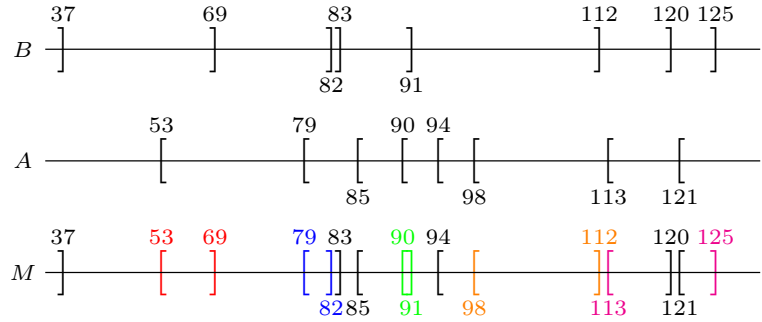

In a matching $M$, a bid or an ask appears at most once. There might be some bids in $B$ which are not matched to any asks in $M$  and some asks in $A$ which are not matched to any bids in $M$. The list of bids present in $M$ is denoted by $B_{M}$ and the list of asks present in $M$ is denoted by $A_M$. For example in Fig.~\ref{fig:matching} the bid with limit price $37$ is not present in $B_M$.

%%%%%%%%%%%%%%TIKz begin %%%%%%%%%%%

%%%%%%%%%%%%%TIKz end    %%%%%%%%%%%%%%%%%%%%

More precisely, for a given list of bids $B$ and list of asks $A$, $M$ is a matching iff, (1) All the bid-ask pairs in $M$ are matchable, (2) $B_M$ is duplicate-free, (3) $A_M$ is duplicate-free, (4) $B_M \subseteq B$, and (5) $A_M \subseteq A$.

\begin{definition}
\textcolor{gray}{matching\_in} $B$ $A$ $M$ := All\_matchable M $\land$ NoDup $B_M$ $\land$ NoDup $A_M$ $\land$ $B_M \subseteq B$ $\land$ $A_M \subseteq A$.
\end{definition}

The term \emph{NoDup $B_M$} in  the above definition indicates that each bid is a request to trade one unit of the item and the items are indivisible.  We use the term $B_M \subseteq B$  to express that each element in the list $B_M$ comes from the list $B$.

Let $B (\geq p)$ represents the bids in $B$ whose limit price is greater than or equal to a given price $p$. In other words, the quantity $|B (\geq p)|$ represents the total demand of the product at a given price $p$ in the market. Similarly, we can use $A  (\leq p)$ to represent all the asks in $A$ whose limit price is less than or equal to the given price $p$.  Hence, the quantity $|A  (\leq p)|$ represents the total supply of the product at the given price $p$.  

Although, in general we can not say much about the relationship between the total demand (i.e. $|B  (\geq p)|$) and supply (i.e. $|A (\leq p)|$) at an arbitrary price $p$, we can certainly prove the following important result about the matched bid ask pairs. 

\begin{lemma}\label{lem:buyersabove}
\textcolor{gray} {buyers\_above\_ge\_sellers}(M: list fill\_type) (B: list Bid) (A: list Ask): $\forall$ p,
  matching\_in $B$ $A$ $M$ $\rw$ $|B_M   (\geq p)|$ $\geq$ $|A_M  (\geq p)|$. 
\end{lemma}  

Lemma \ref{lem:buyersabove} claims that in any valid trade output $M$ and for a given price $p$, the total volume of bids willing to buy at or above the price $p$ is equal to or higher than the total volume of asks willing to sell at a limit price at least $p$. 

Similarly, we prove Lemma \ref{lem:sellersbelow} which states that,  In a matching $M$, the total volume of bids willing to buy at or below a price $p$ is equal to or smaller than the total volume of asks willing to sell at a limit price at most $p$.  

\begin{lemma}\label{lem:sellersbelow}
\textcolor{gray}{sellers\_below\_ge\_buyers} (M: list fill\_type) (B: list Bid) (A: list Ask): $\forall$ p,
  matching\_in $B$ $A$ $M$ $\rw$ $|B_M (\leq p)|$ $\leq$ $|A_M (\leq p)|$.
\end{lemma}  

Additionally, we have the following lemma which provides an upper bound on the cardinality of a matching $M$ using $|B_M  (\geq p)|$  and  $|A_M (\leq p)|$ at a price $p$.
\begin{lemma}\label{lem:buyerright}
\textcolor{gray}{maching\_buyer\_right\_plus\_seller\_left} 
(M: list fill\_type) (B:list Bid) (A:list Ask): $\forall$ p,
(matching\_in $B$ $A$ $M$) $\rw$ $|M|$ $\leq$ $|B_M (\geq p)|$ $+$ $|A_M (\leq p)|$.
\end{lemma}  

It is important to note that the total demand at a certain price $p$ in the market is always greater or equal to the matched demand at a price $p$ or above (i.e. $|B (\geq p)|$ $\geq$ $|B_M (\geq p)|$). Similarly, for total supply at a price $p$ we have $|A (\leq p)|$ $ \geq$ $| A_M (\leq p)|$. These facts when put together with Lemma \ref{lem:buyerright} can help us prove the following result.
\begin{theorem}\label{lem:boundM}
\textcolor{gray}{bound\_on\_M} 
(M: list fill\_type) (B:list Bid) (A:list Ask): $\forall$ p,
(matching\_in $B$ $A$ $M$) $\rw$ $|M|$ $\leq$ $|B  (\geq p)|$ $+$ $|A  (\leq p)|$.
\end{theorem}  
It states that no matching $M$ can achieve a trade volume higher than the sum of the total demand and supply in the market at any given price.

\subsection{Individually Rational Trades}\label{sec:analysis}
An auctioneer assigns a trade price to each matched bid-ask pair. Since the limit price for a buyer is the price above which she does not want to buy, the trade price for this buyer is expected to be below her limit price. Similarly, the trade price for the seller is expected to be above his limit price. Therefore, in any matching it is desired that the trade price of a bid-ask pair lies between their limit prices. A matching which has this property is called an \emph{individual rational (IR)} matching. 
\begin{definition}
\textcolor{gray}{Is\_IR} M := $\forall$ m, m $\in$ M $\rw$ ((bid\_of m) $\ge$ tp m) $\land$ (tp m $\ge$ (ask\_of m)).
\end{definition}

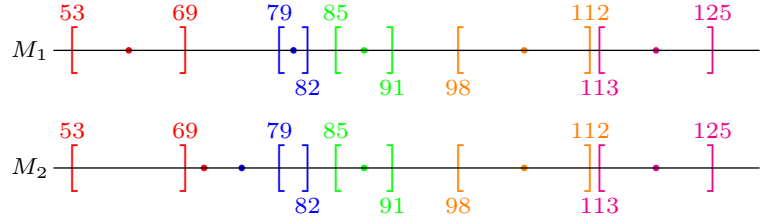
\begin{figure}[h]
\begin{center}
\resizebox {.85\textwidth} {!} {
\begin{tikzpicture}[font=\Large]
%Loop for Lists, top elements, fill colors. 
%Loop for letter A' and M
\foreach \c/\x/\t/\y/\m in {3.3/3.9/red/69/.4, 5.05/5.2/blue/82/-.4, 5.8/6.1/green/91/-.4, 7.5/8.2/orange/112/.4, 8.9/9.5/magenta/125/.4}
{
 \draw [color=\t] (\x, 1.25) node {$]$};
 \draw [color=\t] (\x, 0) node {$]$};

 \draw [color =\t] (\x, 1.25+\m) node {{\scriptsize \y}};
 \draw [color =\t] (\x, \m) node {{\scriptsize \y}};

 \fill[color =\t] (\c,1.25) circle (0.35mm);
}

\foreach \c/\x/\t/\y/\m in {4.1/2.7/red/53/.4,4.5/4.9/blue/79/.4, 5.8/5.5/green/85/.4, 7.5/6.8/orange/98/-.4, 8.9/8.3/magenta/113/-.4}
{
 \draw [color =\t] (\x, 1.25) node {$[$};
 \draw [color =\t] (\x, 0) node {$[$};

 \draw [color =\t] (\x, 1.25+\m) node {{\scriptsize \y}};
 \draw [color =\t] (\x, \m) node {{\scriptsize \y}};
 
 \fill[color =\t] (\c, 0) circle (0.35mm);
}

\draw (2.5,0) -- (10,0);
 
\draw (2.5,1.25) -- (10,1.25);

\draw (2.25,0) node {{\scriptsize $M_2$}};
 
\draw (2.25,1.25) node {{\scriptsize $M_1$}};

\end{tikzpicture}
}
\end{center}
\caption{ The colored dots represent trade prices for matched bid-ask pairs. Matching $M_2$ is not IR but $M_1$ is IR, even though both the matchings contain exactly the same bid-ask pairs.}
\label{fig:IR}
\end{figure}

Note that any matching can be converted to an IR matching without altering its bid-ask pair (See Fig~\ref{fig:IR}). Hence we have the following result,

\begin{theorem}\label{thrm:IR}
\textcolor{gray}{exists\_IR\_matching}: $\forall$ M B A, matching\_in $B$ $A$ $M$ $\rw$ ($\exists$ $M'$, $B_M$ = $B_M'$ $\land$ $A_M$ = $A_M'$ $\land$ matching\_in $B$ $A$ $M'$ $\land$ Is\_IR $M'$).
\end{theorem}

%%%%%%%%%%%%%%TIKz begin %%%%%%%%%%%

%%%%%%%%%%%%%TIKz end    %%%%%%%%%%%%%%%%%%%%

\subsection{Fairness in Competitive Markets}  

 A bid with higher limit price is considered more \emph{competitive} compared to bids with lower limit prices. Similarly, an ask with lower limit price is considered more competitive compared to asks with higher limit prices. In a competitive market, more competitive traders are prioritized for matching. A matching which prioritizes more competitive traders is called a \emph{fair} matching. 

\begin{definition}
\textcolor{gray} {fair\_on\_bids} M B:=
 $\forall$ $b$ $b'$, $b \in B$ $\land$ $b' \in B$  $\rw$ $b>b'$  $\rw$ $b' \in B_M$ $\rw$ $b \in B_M$.
 \end{definition}

\begin{definition}
\textcolor{gray}{fair\_on\_asks} M A:=
 $\forall$ $s$ $s'$, $s \in A$ $\land$ $s' \in A$ $\rw$ $s < s'$ $\rw$ $s' \in A_M$ $\rw$ $s \in A_M$.
\end{definition}

\begin{definition}
\textcolor{gray} {Is\_fair} M B A:= 
fair\_on\_asks M A   $\land$  fair \_on\_bids M B.
\end{definition}

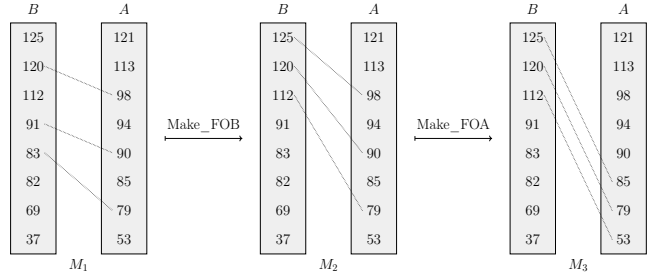
\begin{figure}[h]
\centering
\begin{center}

\resizebox {.7\textwidth} {!} {

\begin{tikzpicture}[font=\Large]

 \draw [fill=lightgray!25] rectangle (22/14,8);

 \foreach \i/\j in {.5/37,1.5/69,2.5/82,3.5/83,4.5/91,5.5/112,6.5/120,7.5/125}
    \path (.5*22/14, \i) node {$\j$};

 \draw [fill=lightgray!25] (2*22/14,0) --(2*22/14,8)-- (3*22/14,8)-- (3*22/14,0)--cycle;

 \foreach \i/\j in {.5/53,1.5/79,2.5/85,3.5/90,4.5/94,5.5/98,6.5/113,7.5/121}
    \path (2.5*22/14, \i) node {$\j$};

 \draw [fill=lightgray!25] (5.5*22/14,0) --(5.5*22/14,8)-- (6.5*22/14,8)-- (6.5*22/14,0)--cycle;

 \foreach \i/\j in {.5/37,1.5/69,2.5/82,3.5/83,4.5/91,5.5/112,6.5/120,7.5/125}
    \path (6*22/14, \i) node {$\j$};

 \draw [fill=lightgray!25] (7.5*22/14,0) --(7.5*22/14,8)-- (8.5*22/14,8)-- (8.5*22/14,0)--cycle;

 \foreach \i/\j in {.5/53,1.5/79,2.5/85,3.5/90,4.5/94,5.5/98,6.5/113,7.5/121}
    \path (8*22/14, \i) node {$\j$};

 \draw [fill=lightgray!25] (11*22/14,0) --(11*22/14,8)-- (12*22/14,8)-- (12*22/14,0)--cycle;

 \foreach \i/\j in {.5/37,1.5/69,2.5/82,3.5/83,4.5/91,5.5/112,6.5/120,7.5/125}
    \path (11.5*22/14, \i) node {$\j$};

 \draw [fill=lightgray!25] (13*22/14,0) --(13*22/14,8)-- (14*22/14,8)-- (14*22/14,0)--cycle;

 \foreach \i/\j in {.5/53,1.5/79,2.5/85,3.5/90,4.5/94,5.5/98,6.5/113,7.5/121}
    \path (13.5*22/14, \i) node {$\j$};

 \foreach \i/\j in {.5/B,2.5/A,6/B,8/A,11.5/B,13.5/A}
    \path (\i*22/14, 8.2) node[above] {$\j$};

%Text Location
  \draw (1.5*22/14, -.7) node[above] {$M_1$};
  \draw (7*22/14, -.7) node[above] {$M_2$};
  \draw (12.5*22/14, -.7) node[above] {$M_3$};

%Curve Arrow: 'out' and 'in' are the angle from x-axis 

%  \draw [-{Latex[length=4.5mm]}] (3.2*22/14,4) to [out=45,in=145] (5.3*22/14,4);

%  \draw [-{Latex[length=4.5mm]}] (8.7*22/14,4) to [out=45,in=145] (10.8*22/14,4);

	\draw[|->] (3.4*22/14,4) -- (5.1*22/14,4); 

	\draw[|->] (8.9*22/14,4) -- (10.6*22/14,4);

%Text Make\_FOB and Make\_FOA

  \draw (4.25*22/14, 4.1) node[above] {Make\_FOB};

  \draw (9.75*22/14, 4.1) node[above] {Make\_FOA};

%Draw dotted $M_1$

  \draw [densely dotted] (.75*22/14,3.5) -- (2.25*22/14,1.5);

  \draw [densely dotted] (.75*22/14,4.5) -- (2.25*22/14,3.5);

  \draw [densely dotted] (.75*22/14,6.5) -- (2.25*22/14,5.5);

%Draw dotted $M_2$

  \draw [densely dotted] (6.25*22/14,5.5) -- (7.75*22/14,1.5);

  \draw [densely dotted] (6.25*22/14,6.5) -- (7.75*22/14,3.5);

  \draw [densely dotted] (6.25*22/14,7.5) -- (7.75*22/14,5.5);

%Draw dotted $M_3$

  \draw [densely dotted] (11.75*22/14,5.5) -- (13.25*22/14,0.5);

  \draw [densely dotted] (11.75*22/14,6.5) -- (13.25*22/14,1.5);

  \draw [densely dotted] (11.75*22/14,7.5) -- (13.25*22/14,2.5);

\end{tikzpicture}

}

\end{center}
\caption{The dotted lines represent  matched bid-ask pairs. The function \emph{make\_FOB} changes $M_1$ into a fair matching on bids $M_2$, whereas \emph{make\_FOA} changes $M_2$ into a fair matching on asks $M_3$.}
\label{fig:fair}
\end{figure}

Here, the predicate \emph{fair\_on\_bids M B}  states that the matching $M$  is fair for the list of buyers $B$. Similarly,  the predicate \emph{fair\_on\_asks M A} states that the matching $M$  is fair for the list of sellers $A$.  A matching which is fair on bids as well as asks is expressed using the predicate \emph{Is\_fair M B A}. Now we can state and prove the following result which states that a fair matching can always be achieved without compromising the cardinality of the matching. 
\begin{theorem}\label{thm:fairMatching}
\textcolor{gray}{exists\_fair\_matching} (Nb: NoDup B) (Na: No Dup A): 
 matching\_in $B$ $A$ $M$ $\rw$ ($\exists$ $M'$,  matching\_in $B$ $A$ $M'$ $\land$ Is\_fair $M'$ $B$ $A$ $\land$ $|M| = |M'|$).
\end{theorem}
\noindent \emph{Proof Idea.} We prove this statement  by converting a matching into a fair matching without changing its cardinality. In order to achieve this we use functions \emph{make\_FOB} and \emph{make\_FOA} (See Fig~\ref{fig:fair}). The function \emph{make\_FOB} produces a matching which is fair on bids from an input matching M and a list of bids B both of which are sorted in decreasing order of their bid prices (Lemma \ref{lem:fob}). Moreover, since \emph{make\_FOB} does not change any of the asks in M, it results in a matching of size $|M|$.  Once we get a fair matching on bids, we use a similar function \emph{make\_FOA} to produce a matching which is fair on the asks. Finally, the correctness proofs of  \emph{make\_FOB} and  \emph{make\_FOA} can be composed to complete the proof of the present theorem. $\square$

%%%%%%%%%%%%%%%%%%%%%%%

The functions \emph{make\_FOB} and  \emph{make\_FOA} are both recursive in nature and have identical definitions. Therefore, it is sufficient 
to discuss the properties of \emph{make\_FOB} which is defined recursively as follows.
\begin{Verbatim}[fontsize=\footnotesize]
  Fixpoint Make_FOB (M) (B):= match (M,B) with 
	|(nil,_) => nil
	|(m::M',nil) => nil
	|(m::M',b::B') =>  (Mk_fill b (ask_of m) (tp m))::(Make_FOB M' B')
	end.
\end{Verbatim}

 In each step the function \emph{make\_FOB} picks the top bid-ask pair, say $(b,a)$ in $M_1$ and replaces $b$ with the most competitive bid available in $B$,  resulting in a matching $M_2$ (See Fig.~\ref{fig:fair}). Note that \emph{make\_FOB} does not change any of the asks in M. Moreover, due to the recursive nature of \emph{make\_FOB} on B, a bid is not repeated in the process of replacement (i.e., $B_{M_2}$ is duplicate-free). Therefore, we would like to have the following lemma. 
 \begin{lemma}\label{lem:fobcorrect}
$\forall$ $M$ $B$,  (Sorted $\downarrow_{bp}$ $M$) $\rw$ (Sorted $\downarrow_{bp}$ $B$) $\rw$  matching\_in $B$ $A$ $M$ $\rw$  fair\_on\_bids (Make\_FOB $M$ $B$) $B$. 
\end{lemma}
\noindent \emph{Induction Failure}: The function \emph{make\_FOB} is recursive on both $B$ and $M$, and hence a proof of this lemma is expected using an inductive argument on the structure of $B$. Although the theorem is true, an attempt to prove it using induction on B will fail. Let  $M = (b_1, a_1)::(b_2, a_2)::M''$ and $B = b_2::b_1::B''$, where both $B$ and $M$ are sorted by decreasing bid prices and $(bp\  b_1) = (bp \  b_2)$. After the first iteration, the \emph{make\_FOB} will calls itself on $M' = (b_2, a_2)::M''$ and  $B' = (b_1::B'')$. In the inductive proof, in order to use the induction hypothesis we need to prove that $M'$ is a matching for the list of bids in $B'$. This is clearly not true since $B_{M'}$ is not a subset of $B'$ since $b_2 \notin B'$ but $b_2 \in B_{M'}$. This complication arises because we are dealing with all the information contained in the bids while the proof requires reasoning only based on their limit prices. We resolve this difficulty by systematically mapping the properties of $M$, $B$ and $A$ to the properties of their corresponding price columns. For example, we have the following result on the prices of $B$.

\begin{lemma}\label{lem:projectsub} 
\textcolor{gray}{sorted\_nodup\_is\_sublistB}: $\forall$ $B_1$ $B_2$,   NoDup $B_1$ $\rw$ NoDup $B_2$ $\rw$ Sorted $\downarrow_{bp}$ $B_1$ $\rw$ Sorted $\downarrow_{bp}$ $B_2$ $\rw$  $B_1 \subset B_2$ $\rw$ sublist $P_{B_1}$ $P_{B_2}$. 
\end{lemma} 

Here, $P_B$ is projection of the limit prices of bids in $B$. The term (\emph{sublist} $P_{B_1}$ $P_{B_2}$) represents the sub-sequence relation between the lists $P_{B_1}$ and $P_{B_2}$. Furthermore, we have the following lemmas specifying the sub-list relation between lists.

\begin{lemma} 
\textcolor{gray}{sublist\_intro1}: $\forall$ a, sublist l s $\rw$ sublist l (a::s).
\end{lemma}

\begin{lemma}\label{lem:sublistInd} 
\textcolor{gray}{sublist\_elim3a}: $\forall$ a e, sublist (a::l) (e::s) $\rw$ sublist l s.
\end{lemma}

Note the recursive nature of the \emph{sublist} relation on both its arguments, as evident in Lemma~\ref{lem:sublistInd}. It makes inductive reasoning feasible for the statements where \emph{sublist} is in the antecedent. Hence, we use the sublist relation to state and prove the following result.

\begin{lemma}\label{lem:fob}
\textcolor{gray} {mfob\_fair\_on\_bid} M B:
  (Sorted $\downarrow_{bp}$ $M$) $\rw$ (Sorted $\downarrow_{bp}$ $B$) $\rw$  sublist $P_{B_M}$  $P_B$ $\rw$  fair\_on\_bids (Make\_FOB $M$ $B$)  $B$. 
\end{lemma}

Similarly, we can state and prove the following result which specifies the function \emph{make\_FOA}.

\begin{lemma}\label{lem:foa}
\textcolor{gray} {mfob\_fair\_on\_ask} M A: 
 (Sorted $\uparrow_{sp}$ $M$) $\rw$ (Sorted $\uparrow_{sp}$ $A$) $\rw$ sublist $P_{A_M}$ $P_A$  $\rw$
  fair\_on\_asks (Make\_FOA $M$ $A$) $A$. 
\end{lemma}

Since the fair matching is obtained by composing the functions \emph{Make\_FOA} and \emph{Make\_FOB},  we can combine the proofs of Lemma \ref{lem:foa} and Lemma \ref{lem:fob} to obtain the complete proof of Theorem~\ref{thm:fairMatching}. 

\subsection{Liquidity and Perceived-fairness in the Markets} \label{sec:liqFair}
The liquidity in any market is a measure of how quickly one can trade in the market without much cost. One way to increase the liquidity is to maximize the number of matched bid-ask pairs.  For a given list of bids $B$ and list of asks $A$ we say a matching $M$ is  a maximum matching if no other matching $M'$ on the same $B$ and $A$ contains more matched bid-ask pairs than $M$. 

\begin{definition}
\textcolor{gray}{Is\_MM} $M$ $B$ $A$ := (matching\_in $B$ $A$ $M$) $\land$ 
($\forall$ $M'$, matching\_in $B$ $A$ $M'$ $\rightarrow$ $|M'| \leq |M|$).
\end{definition}

\begin{figure}[h]
\begin{center}

\resizebox {.6\textwidth} {!} {

\begin{tikzpicture}[font=\Large]

 \draw [fill=gray!25] rectangle (22/14,2);

 \draw (.5*22/14, .5) node {$80$};

 \draw (.5*22/14, 1.5) node {$100$};

 \draw [fill=lightgray!25] (2*22/14,0) --(2*22/14,2)-- (3*22/14,2)-- (3*22/14,0)--cycle;

 \draw (2.5*22/14, .5) node {$90$};

 \draw (2.5*22/14, 1.5) node {$70$};

 \draw [fill=lightgray!25] (6*22/14,0) --(6*22/14,2)-- (7*22/14,2)-- (7*22/14,0)--cycle;

 \draw (6.5*22/14, .5) node {$80$};

 \draw (6.5*22/14, 1.5) node {$100$};

 \draw [fill=lightgray!25] (8*22/14,0) --(8*22/14,2)-- (9*22/14,2)-- (9*22/14,0)--cycle;

 \draw (8.5*22/14, .5) node {$90$};

 \draw (8.5*22/14, 1.5) node {$70$};

 \path (.5*22/14, 2.2) node[above] {$B$};

 \path (2.5*22/14, 2.2) node[above] {$A$};

 \path (6.5*22/14, 2.2) node[above] {$B$};

 \path (8.5*22/14, 2.2) node[above] {$A$};

%Text Location
  \draw (1.5*22/14, -1) node[above] {(a) $UM$};
  \draw (7.5*22/14, -1) node[above] {(b) $MM$};

%Draw dotted $produce\_MM$

  \draw [densely dotted] (0.75*22/14,1.5) -- (2.25*22/14,1.5);

  \draw [densely dotted] (6.75*22/14,1.5) -- (8.25*22/14,0.5);

  \draw [densely dotted] (6.75*22/14,0.5) -- (8.25*22/14,1.5);

\end{tikzpicture}
}
\end{center}
\caption{The only individually rational matching of size two is not uniform.}
\label{fig:mmum}

\end{figure}
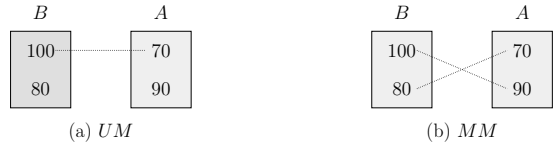

Designing a mechanism for a maximum matching is an important aspect of a double sided auction. In certain situations, to produce a maximum matching, bid-ask pairs must be assigned different trade prices (Fig.~\ref{fig:mmum}). However, different prices simultaneously for the same product leads to dissatisfaction amongst some of the traders. A mechanism which clears all the matched bid-ask pairs at a single trade price is called a \emph{uniform matching} (or \emph{perceived-fairness}). 

%%%%%%%%%%%%%%%TIKZ begin%%%%%%%%%%%%%%%%%%

%%%%%%%%%%%%%%%TIKZ end%%%%%%%%%%%%%%%%%%%%%

\section{Optimizing Trades in Financial Markets}\label{sec:optimization}
In  Section \ref{sec:liqFair}, we observed that a maximum matching may not be a uniform matching. In this Section, we present two broad classes of double sided auction mechanisms: a maximum matching mechanism and a uniform price mechanism. While the maximum matching mechanism tries to maximize the overall volume of the trade, the uniform price mechanism tries to obtain a uniform matching of maximum possible cardinality.  
\subsection{A Maximum Maching Mechanism} 
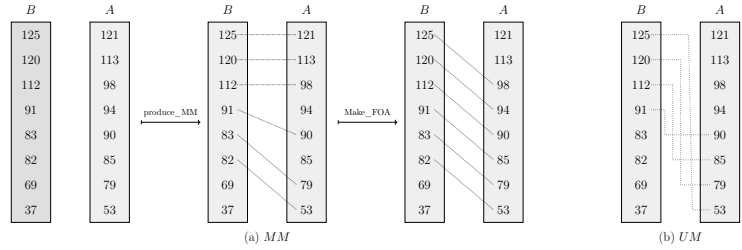
\begin{figure}[h]
\centering
\begin{center}

\resizebox {.8\textwidth} {!} {

\begin{tikzpicture}[font=\Large]

 \draw [fill=gray!25] rectangle (22/14,8);

 \foreach \i/\j in {.5/37,1.5/69,2.5/82,3.5/83,4.5/91,5.5/112,6.5/120,7.5/125}
    \path (.5*22/14, \i) node {$\j$};

 \draw [fill=lightgray!25] (2*22/14,0) --(2*22/14,8)-- (3*22/14,8)-- (3*22/14,0)--cycle;

 \foreach \i/\j in {.5/53,1.5/79,2.5/85,3.5/90,4.5/94,5.5/98,6.5/113,7.5/121}
    \path (2.5*22/14, \i) node {$\j$};

 \draw [fill=lightgray!25] (5*22/14,0) --(5*22/14,8)-- (6*22/14,8)-- (6*22/14,0)--cycle;

 \foreach \i/\j in {.5/37,1.5/69,2.5/82,3.5/83,4.5/91,5.5/112,6.5/120,7.5/125}
    \path (5.5*22/14, \i) node {$\j$};

 \draw [fill=lightgray!25] (7*22/14,0) --(7*22/14,8)-- (8*22/14,8)-- (8*22/14,0)--cycle;

 \foreach \i/\j in {.5/53,1.5/79,2.5/85,3.5/90,4.5/94,5.5/98,6.5/113,7.5/121}
    \path (7.5*22/14, \i) node {$\j$};

 \draw [fill=lightgray!25] (10*22/14,0) --(10*22/14,8)-- (11*22/14,8)-- (11*22/14,0)--cycle;

 \foreach \i/\j in {.5/37,1.5/69,2.5/82,3.5/83,4.5/91,5.5/112,6.5/120,7.5/125}
    \path (10.5*22/14, \i) node {$\j$};

 \draw [fill=lightgray!25] (12*22/14,0) --(12*22/14,8)-- (13*22/14,8)-- (13*22/14,0)--cycle;

 \foreach \i/\j in {.5/53,1.5/79,2.5/85,3.5/90,4.5/94,5.5/98,6.5/113,7.5/121}
    \path (12.5*22/14, \i) node {$\j$};

 \draw [fill=lightgray!25] (15.5*22/14,0) --(15.5*22/14,8)-- (16.5*22/14,8)-- (16.5*22/14,0)--cycle;

 \foreach \i/\j in {.5/37,1.5/69,2.5/82,3.5/83,4.5/91,5.5/112,6.5/120,7.5/125}
    \path (16*22/14, \i) node {$\j$};

 \draw [fill=lightgray!25] (17.5*22/14,0) --(17.5*22/14,8)-- (18.5*22/14,8)-- (18.5*22/14,0)--cycle;

 \foreach \i/\j in {.5/53,1.5/79,2.5/85,3.5/90,4.5/94,5.5/98,6.5/113,7.5/121}
    \path (18*22/14, \i) node {$\j$};

 \foreach \i/\j in {.5/B,2.5/A,5.5/B,7.5/A,10.5/B,12.5/A,16/B,18/A}
    \path (\i*22/14, 8.2) node[above] {$\j$};

%Text Location
  \draw (6.5*22/14, -1) node[above] {(a) $MM$};
  \draw (17*22/14, -1) node[above] {(b) $UM$};

%Curve Arrow: 'out' and 'in' are the angle from x-axis 

%  \draw [-{Latex[length=4.5mm]}] (3.2*22/14,4) to [out=45,in=145] (5.3*22/14,4);

%  \draw [-{Latex[length=4.5mm]}] (8.7*22/14,4) to [out=45,in=145] (10.8*22/14,4);

	\draw[|->] (3.3*22/14,4) -- (4.8*22/14,4); 

	\draw[|->] (8.3*22/14,4) -- (9.8*22/14,4);

%Text Make\_FOB and Make\_FOA

  \draw (4.05*22/14, 4.1) node[above] {{\normalsize produce\_MM}};

  \draw (9.05*22/14, 4.1) node[above] {{\normalsize Make\_FOA}};

%Draw dotted $produce\_MM$

  \draw [densely dotted] (5.75*22/14,2.5) -- (7.25*22/14,0.5);

  \draw [densely dotted] (5.75*22/14,3.5) -- (7.25*22/14,1.5);

  \draw [densely dotted] (5.75*22/14,4.5) -- (7.25*22/14,3.5);

  \draw [densely dotted] (5.75*22/14,5.5) -- (7.25*22/14,5.5);

  \draw [densely dotted] (5.75*22/14,6.5) -- (7.25*22/14,6.5);

  \draw [densely dotted] (5.75*22/14,7.5) -- (7.25*22/14,7.5);

%Draw dotted $make\_FOA$

  \draw [densely dotted] (10.75*22/14,2.5) -- (12.25*22/14,0.5);

  \draw [densely dotted] (10.75*22/14,3.5) -- (12.25*22/14,1.5);

  \draw [densely dotted] (10.75*22/14,4.5) -- (12.25*22/14,2.5);

  \draw [densely dotted] (10.75*22/14,5.5) -- (12.25*22/14,3.5);

  \draw [densely dotted] (10.75*22/14,6.5) -- (12.25*22/14,4.5);

  \draw [densely dotted] (10.75*22/14,7.5) -- (12.25*22/14,5.5);

%Draw dotted $UM$

  \draw [dotted] (16.25*22/14,4.5) -- (16.6*22/14,4.5) -- (16.6*22/14,3.5) -- (17.75*22/14,3.5);

  \draw [dotted] (16.25*22/14,5.5) -- (16.8*22/14,5.5) -- (16.8*22/14,2.5) -- (17.75*22/14,2.5);

  \draw [dotted] (16.25*22/14,6.5) -- (17*22/14,6.5) -- (17*22/14,1.5) -- (17.75*22/14,1.5);
  
  \draw [dotted] (16.25*22/14,7.5) -- (17.20*22/14,7.5) -- (17.30*22/14,0.5) -- (17.75*22/14,0.5);

\end{tikzpicture}

}

\end{center}
\caption{(a) At each iteration \emph{produce\_MM} selects a most competitive available bid and then pairs it with the largest matchable ask. The output of this function is already fair on bids. In the second step, the function \emph{make\_FOA} converts this output into fair matching. (b) Maximum matching amongst uniform. Note that, the size of both the matchings are different. }
\label{fig:mm}
\end{figure}

We will now discuss a matching mechanism which produces maximum trade volume while maintaining the  fairness criterion. This scheme produces the same output as the one proposed in \cite{NiuP13}. However, there are some important differences in both mechanisms. The algorithm suggested in \cite{NiuP13} is a non recursive method which generates the final trade in two steps; the algorithm first determines the cardinality $n$ of a maximum matching on the given set of bids and asks and  then in the next step it produces a fair matching of cardinality $n$. On the other hand, we use a  recursive function \emph{produce\_MM} on the lists of bids and asks to produce a maximum matching which is then converted into a fair matching using the already defined function \emph{make\_FOA} (See Fig.~\ref{fig:mm}(a)). We follow this approach because it allows us to easily compose the correctness proof of these individual functions to validate the properties of the final trade generated by the whole mechanism. 

%%%%%%%%%%%%%%%%%%%%%TIKZ image%%%%%%%%%%%%%%%%%%%%%%%%%%%%%%%%%%%%%%%%%

%%%%%%%%%%%%%%%%%%%%%End TikZ Image%%%%%%%%%%%%%%%%%%%%%%%%%%%%%%%%%%%%%

\vspace{-.5cm}
\begin{Verbatim}[fontsize=\footnotesize]
Fixpoint produce_MM (B) (A) := match (B, A) with
  |(nil, _) => nil
  |(b::B', nil) => nil              
  |(b::B', a::A') =>  match (a <= b) with
     |true => {|bid_of:=b; ask_of:=a; tp:=(bp b)|}::(produce_MM B' A')
     |false => produce_MM B A'
    end
  end. 
\end{Verbatim}

The correctness proof of \emph{produce\_MM} is obtained using an inductive argument on the structure of the input lists. At each iteration  \emph{produce\_MM} generates a matchable bid-ask pair (See Fig.~\ref{fig:mm}(a)). Due to the recursive nature of  function  \emph{produce\_MM} on both $B$ and $A$, it never pairs any bid with more than one ask. This ensures that the list of bids in matching (i.e. $B_M$) is duplicate-free. Note that \emph{produce\_MM} tries to match a bid until it finds a matchable ask. The function terminates when either all the bids are matched or it encounters a bid for which no matchable ask is available. The following theorem states that the function \emph{produce\_MM} produces a maximum matching when both  $B$ and $A$ are sorted in a decreasing order of the limit prices.
\begin{theorem}\label{thm:maxMatching}
\textcolor{gray} {produce\_MM\_is\_MM} (Nb: NoDup B) (Na: NoDup A): Sorted $\downarrow_{bp}$ $B$ $\rw$ Sorted $\downarrow_{sp}$ $A$ $\rw$ Is\_MM (produce\_MM $B$ $A$) $B$ $A$.
\end{theorem}
The proof idea for the above theorem has been moved to Appendix~\ref{proof:mm}.

Now that we  proved the maximality property of \emph{produce\_MM}  we can produce a fair as well as maximum matching by applying the functions \emph{Make\_FOA}  and \emph{Make\_FOB} to the output of \emph{produce\_MM}. More precisely, for a given list of bids $B$ and list of asks $A$, we have  the following result stating that there exists a matching which is both maximum and fair. 

\begin{theorem}\label{thm:existfairMM}
\textcolor{gray}{exists\_fair\_maximum} (B: list Bid)(A: list Ask): $\exists$ $M$, (Is\_fair $M$ $B$ $A$ $\land$ Is\_MM $M$ $B$ $A$).
\end{theorem}

\subsection{Trading at Equilibrium Price}\label{sec:matchingInMarkets}

An important aspect of the opening session of a market is to discover a single price (equilibrium price) at which maximum demand and supply can be matched. Most exchanges execute trade during this session at an equilibrium price. An equilibrium price determined at exchanges is usually the limit price of a bid or ask from a bid-ask pair such that the uniform matching produced in this session remains individual rational. We will now describe a function \emph{produce\_UM} which produces an individually rational matching which is fair and maximum among all uniform matchings.

\begin{Verbatim}[fontsize=\footnotesize]
Fixpoint pair_uniform (B:list Bid) (A:list Ask):=  match (B,A) with
  |(nil, _) => nil
  |(_,nil)=> nil
  |(b::B',a::A') => match (a <= b) with
  	|false => nil
  	|true  =>{|bid_of:= b;ask_of:= a; tp:=(bp b)|}::pair_uniform B' A'
  	end
	end.
Definition uniform_price B A := bp (bid_of (last (pair_uniform B A))).
Definition produce_UM B A:= 
replace_column (pair_uniform B A) (uniform_price B A).
\end{Verbatim}

The function \emph{pair\_uniform} output bid-ask pairs, \emph{uniform\_price} computes the uniform price and finally \emph{produce\_UM} produces a uniform matching. The function \emph{pair\_uniform} is recursive and  matches the largest available bid in $B$ with the smallest available ask in $A$ at each iteration (See Fig.~\ref{fig:mm}(b)). This function terminates when the most competitive bid available in $B$ is not matchable with any available ask in $A$. 

The following theorem states that the function \emph{produce\_UM} produces a maximum matching among all uniform matchings when the list of bids $B$ is sorted in a decreasing order of the limit prices and the list of asks $A$ is sorted in an increasing order of the limit prices.

\begin{theorem}\label{thm:uniformMM}
\textcolor{gray}{UM\_is\_maximal\_Uniform} (B: list Bid) (A:list Ask): Sorted $\downarrow_{bp}$  $B$ $\rw$ Sorted $\uparrow_{sp}$ $A$ $\rw$ $\forall$ $M$, Is\_uniform $M$ $\rw$ $|M| \leq |produce\_UM \ B \ A |$.
\end{theorem}

The proof idea for the above theorem has been moved to Appendix~\ref{proof:um}.

\iffalse
\noindent \emph{Proof Idea}: Let $M$ be any arbitrary IR and uniform matching on the list of bids $B$ and list of asks $A$ where each matched bid-ask pair is traded at price $t$. We need to prove that $m \leq$ |\emph{(produce\_UM $B$ $A$)}| where $m$ is the number of matched bid-ask pairs in the matching $M$. Observe that in any individually rational and uniform matching the number of bids above  the trade price is same as the number of asks below the trade price. Therefore, there are at least $m$ bids above $t$ and $m$ asks below $t$ in $B$ and $A$ respectively. Since at each step the function \emph{pair\_uniform} pairs the largest bid available in $B$ with the smallest ask available in $A$ it must produce  at least $m$ bid-ask pairs. Hence for the list of bids $B$ and list of asks $A$ the function \emph{produce\_UM} produces a uniform matching which is of size at least $m$.
\fi

Next, we prove that the \emph{produce\_UM} generates a maximum matching among all uniform matchings which is also fair when the list of bids $B$ is sorted in a decreasing order of the limit prices and the list of asks $A$ is sorted in an increasing order of the limit prices. In order to prove this, we first prove the following two lemmas.

\begin{lemma}\label{thm:uniformfairask}
\textcolor{gray}{UM\_pair\_fair\_on\_asks} (B: list Bid) (A:list Ask): Sorted $\downarrow_{bp}$  $B$ $\rw$ Sorted $\uparrow_{sp}$ $A$ $\rw$
fair\_on\_asks (pair\_uniform $B$ $A$) $A$.
\end{lemma}

\begin{lemma}\label{thm:uniformfairbid}
\textcolor{gray}{UM\_pair\_fair\_on\_bids} (B: list Bid) (A:list Ask): Sorted $\downarrow_{bp}$  $B$ $\rw$ Sorted $\uparrow_{sp}$ $A$ $\rw$
fair\_on\_bids (pair\_uniform $B$ $A$) $B$.
\end{lemma}

\begin{theorem}\label{thm:uniformfair}
\textcolor{gray}{UM\_fair} (B: list Bid) (A:list Ask)(m:fill\_type): Sorted $\downarrow_{bp}$  $B$ $\rw$ Sorted $\uparrow_{sp}$ $A$ $\rw$
Is\_fair (produce\_UM $B$ $A$) $B$ $A$.
\end{theorem}

The proof of Theorem~\ref{thm:uniformfair} is similar to the proof of Theorem~\ref{thm:fairMatching} once we use Lemmas~\ref{thm:uniformfairask} and \ref{thm:uniformfairbid}.

\section{Conclusion and Future Works}\label{sec:conclusion}

In this work, we developed a formal framework to verify important properties of matching algorithms used by the exchanges. These algorithms use double sided auctions to match multiple buyers with multiple sellers during different sessions of trading. We presented correctness proofs for two important classes of double sided auction mechanisms: uniform price algorithms and maximum matching algorithms. 

An important direction of future work is the individual analysis of various orders types which are important for the continuous markets (e.g. limit orders, market orders, stop-loss orders, iceberg orders, fill or kill (FOK), immediate or cancel (IOC) etc.). This would require maintaining a priority queue based on the various  attributes of these orders. A formal analysis of these order attributes together with the verification of trading mechanisms can provide a formal foundation which will be useful in the rigorous analysis of other market behaviors at large. Also for continuous markets, due to the various order types, it becomes important to consider multiple unit orders which requires more work. Moreover, the insights gained from these attempts to formalize the overall trading mechanism can be helpful in developing robust as well as efficient trading systems of the future which can be used directly at the exchanges. 

\section*{Acknowledgment} We thanks N. Raja and Mohit Garg for many useful suggestions and discussions that has improved this work. 

\bibliographystyle{plain} 
%% Bibliography
\bibliography{auction}

\newpage

\appendix\label{appendix}
\section{Appendix}
\subsection{Proof idea for Theorem 5.}
\label{proof:mm}
\begin{appendixthm}\label{thm:maxMatchingproof}
\textcolor{gray} {produce\_MM\_is\_MM} (Nb: NoDup B) (Na: NoDup A): Sorted $\downarrow_{bp}$ B $\rw$ Sorted $\downarrow_{sp}$ A $\rw$ Is\_MM (produce\_MM B A) B A.
\end{appendixthm}

%%%%%%%%%%%%%%%%%%%%%%%%%%%%%%%%%%%
\noindent \emph{Proof Idea}: We prove this result using induction on the cardinality of list $A$. Let $M$ be an arbitrary matching on the list of bids $B$ and list of asks $A$. Moreover, assume that $b$ and $a$ are the topmost bid and ask present in $B$ and $A$, respectively (i.e. $A = (a::A')$ and $B = (b::B')$). We prove $|M| \leq$ |\emph{produce\_MM} $B$ $A$| in  the following two cases.

\noindent \emph{Case-1} ($b<a$): In this case the function \emph{produce\_MM} computes a matching on $B$ and $A'$. Note that due to the induction hypotheses (i.e. \tw{IH}) this is a maximum matching for $B$ and $A'$. Since the limit price of ask $a$ is more than the most competitive bid $b$ in $B$, it cannot be present in any matching of $B$ and $A$. Therefore a maximum matching on $B$ and $A'$ is also a maximum matching on $B$ and $A$. Hence we have |$M$| $\leq$ | \emph{produce\_MM} $B$ $A$ |.

\begin{figure}[h]
\begin{center}
\resizebox {.9\textwidth} {!} {
\begin{tikzpicture}[font=\Large]
%Loop for Lists, top elements, fill colors. 
 \foreach \x/\t/\y in {0/b/0,2/a/0,5/b/0,7/a/0,11.5/b/0,13.5/a/0,16.5/b/0,18.5/a/0,0/b/7,2/a/7,6/b/7,8/a/7,12/b/7,14/a/7,17/b/7,19/a/7}
{
 \draw[draw=none, fill=lightgray] (\x,\y) --(\x,\y+4)-- (\x+1,\y+4)-- (\x+1,\y)--cycle;
 \draw (\x,\y) --(\x,\y+5)-- (\x+1,\y+5)-- (\x+1,\y)--cycle; 
 \draw (\x+.5, \y+4.5) node {$\t$};
}
%Loop for letter B'
\foreach \x/\t/\y in {0/B/0,5/B/0,11.5/B/0,16.5/B/0,0/B/7,6/B/7,12/B/7,17/B/7}
{
 \draw (\x-.5, \y+2) node {$\t^{'}$};
}
%Loop for letter A' and M
\foreach \x/\t/\y/\m in {2/A/0/M,7/A/0/M_1,13.5/A/0/M,18.5/A/0/M_1,2/A/7/M,8/A/7/M,14/A/7/M,19/A/7/M_1}
{
 \draw (\x+1.5, \y+2) node {$\t^{'}$};
 \draw (\x - .5 ,\y) node[above] {$\m$};
}
%Loop for dash lines
\foreach \a/\b/\c/\d in {0/4/2/2,5/4/7/4,11.5/2/13.5/4,16.5/4/18.5/4, 0/11/2/11, 12/11/14/9,12/9/14/11,17/9/19/9,17/11/19/11}
{
 \draw [dashed] (\a +.75, \b +.5) -- (\c +.25, \d +.5);
}
%Loop for Letter a' b'
\foreach \x/\t/\y in {2/a/0,7/a/0,11.5/b/0,16.5/b/0,12/b/7,14/a/7,17/b/7,19/a/7}
{
 \draw (\x +.5, \y+2.5) node {$\t^{'}$};
}
%Loop for other
\foreach \x/\t/\y in {2/d/0, 13.5/e/0, 14/c/7}
{
\draw (\x+2, \y-1) node {(\t)}; 
\draw (\x+2, \y+2.75) node {$\implies$};
}

\draw (1.5, 6) node {(a)}; 
\draw (6.5, 6) node {(b)}; 

\end{tikzpicture}
}
\end{center}
\caption{Five subcases of Case-2. The dotted line shows a matched bid-ask pair in $M$.  Both $B$ and $A$ are sorted in decreasing order of their limit prices. }
\label{fig:mmProof}
\end{figure}
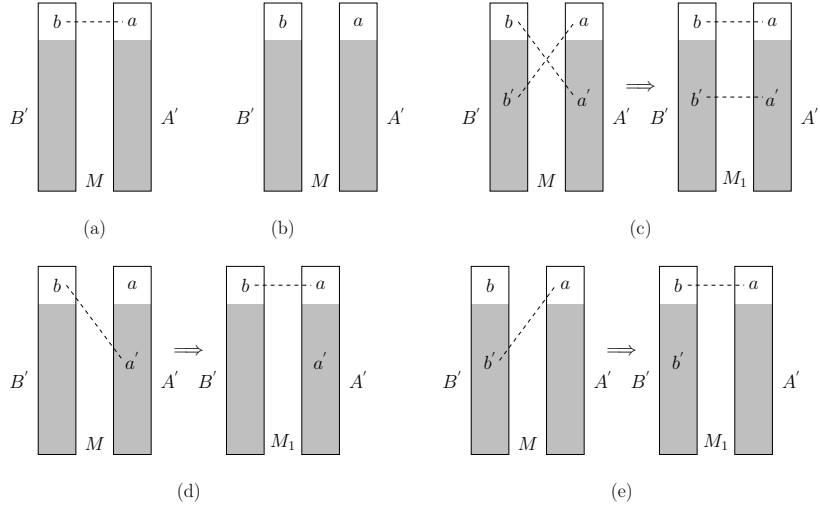

\noindent \emph{Case-2} ($a \leq b$): In this case \emph{produce\_MM} produces a matching of cardinality $m + 1$ where $m$ is the cardinality of matching \emph{produce\_MM $B'$ $A'$}. We need to prove that $|M| \leq  m+1$. Note that due to induction hypothesis  the matching \emph{produce\_MM $B'$ $A'$} is a maximum matching on $B'$ and $A'$.  Hence no matching on $B'$ and $A'$ can have cardinality bigger than $m$. Without loss of generality, we can assume that $M$ is also sorted in a decreasing order of bid prices. Now we further split this case into the following five subcases (see Fig~\ref{fig:mmProof}).

$\triangleright$ \emph{C2-A} ($M = (b,a)::M'$) : In this case  bid $b$ is matched to ask $a$ in the matching $M$ (see Fig~\ref{fig:mmProof} (a)). Note that $M'$ is a matching on $B'$ and $A'$. Since $|M'| \leq m$, we have $|M|= |M'|+1 \leq m+1$.

$\triangleright$ \emph{C2-B}  ($b \notin B_M  \land a \notin A_M$) : In this case neither bid $b$ nor ask $a$ is present in matching $M$ (see Fig~\ref{fig:mmProof} (b)). Therefore $M$ is a matching on $B'$ and $A'$. Hence we have $|M| \leq m < m+1$.

 $\triangleright$ \emph{C2-C}  $(b,a') \in M \land (b',a) \in M$ : In this case we can obtain another matching $M_1$ of the same cardinality as $M$ (see Fig~\ref{fig:mmProof} (c)) where $(b,a) \in M_1$ and $(b',a') \in M_1$. Note that all other entries of $M_1$ is same as $M$. Therefore we have $M_1 = (b,a)::M'$ where $M'$ is a matching on $B'$ and $A'$. Since $|M'| \leq m$, we have $|M|=|M_1| \leq m+1$.

 $\triangleright$ \emph{C2-D}  $(b,a')\in M \land a \notin A_M$ : In this case we can obtain another matching $M_1$ of same cardinality as $M$ (see Fig~\ref{fig:mmProof} (d)) where $(b,a) \in M_1$. 
Therefore we have $M_1 = (b,a)::M'$ where $M'$ is a matching on $B'$ and $A'$.
Since $|M'| \leq m$, we have $|M|=|M_1| \leq m+1$.

$\triangleright$ \emph{C2-E}  $(b',a)\in M \land b \notin B_M$ : In this case we can obtain another matching $M_1$ of same cardinality as $M$ (see Fig~\ref{fig:mmProof} (e)) where $(b,a) \in M_1$. Therefore we have $M_1 = (b,a)::M'$ where $M'$ is a matching on $B'$ and $A'$. Since $|M'| \leq m$, we have $|M|=|M_1| \leq m+1$. $\square$ 

Note that all the cases in the above proof correspond to  predicates which can be expressed using  only the membership predicate on lists. Since we have  decidable equality on the elements of the lists all these predicates are also decidable. Hence, we can do case analysis on them without assuming any axiom. 

%%%%%%%%%%%%%%TIKz begin %%%%%%%%%%%

%%%%%%%%%%%%%TIKz end    %%%%%%%%%%%%%%%%%%%%

\subsection{Proof idea for Theorem 6.}
\label{proof:um}

\begin{appendixthm}
\textcolor{gray}{UM\_is\_maximal\_Uniform} (B: list Bid) (A:list Ask): Sorted $\downarrow_{bp}$  $B$ $\rw$ Sorted $\uparrow_{sp}$ $A$ $\rw$ $\forall$ $M$, Is\_uniform $M$ $\rw$ $|M| \leq |(produce\_UM\, B A )|$.
\end{appendixthm}

The proof of this Theorem is very similar to the proof of Theorem~\ref{thm:maxMatchingproof}. We prove it using induction on $A$. Observe that, if $M$ is a uniform matching then every bid $b\in B_M$ is matchable with every other ask $a \in A_M$. 

Let $M$ be an arbitrary matching on $B$ and $A$ which is uniform and individual rational and the uniform trade price is $p$. It is suffice to prove that $|M|\le|(produce\_UM \, B\, A)|$. With the induction on $A$, we get the following cases.

\begin{enumerate}[label=\Roman*]
\item {\boldmath$A=nil \lor B=nil$:} In this case: $|M|=|(produce\_UM \,  B\, A)|=0$.

\item {\boldmath$A=a::A' \land B=b::B'$:} In this case either $b<a$ or $b\ge a$. When $b<a$, we can prove that $|M|=0$ and $|(produce\_UM\, B\, A)|=0$. It is easy to see that if the largest bid ($b$) is not matchable with the smallest ask ($a$) then no other bid can match with any other ask and the size of any matching in this case is zero. When $b\ge a$ there are three possibilities. (a) Either both $b$ and $a$ appears in the bid-ask pairs of $M$, (b) Only one of the $b$ or $a$ appears in the bids or asks of $M$ and (c) When both of them does not appears in the $B_M$ or $A_M$. 

Note that, since we are doing induction on $A$, we get the following induction hypothesis. 

\begin{equation}
 \forall B, \, Is\_uniform \, M'\, B\, A' \rw (| M' |) \le | (produce\_UM\, B\, A') | 
\end{equation}

Now we need to prove that $|M|\le|(produce\_UM\,  B\, A)|$ where $M$ is a uniform and IR matching in $B$ and $A$. Observe that $(produce\_UM \, B\, A)$ will pair $b$ with $a$ in first iteration so 

\begin{equation}
|(produce\_UM \, B\, A)| = |(produce\_UM \, B'\, A')| + 1
\end{equation}

\begin{enumerate}[label=(\alph*)]
\item Both $b$ and $a$ appears in bids and asks of $M$: Let $m_1,m_2 \in M$ such that $a=ask\_of \, m_1$ and $b=bid\_of \, m_2$. Now consider a matching $M''$ such that $M''=(b,a,p)::(bid\_of \, m_1,ask\_of \, m_2,p)::(M \setminus \{m_1,m_2\})$. It is easy to see $M'=(bid\_of \, m_1,ask\_of \, m_2,p)::(M \setminus \{m_1,m_2\})$ is a uniform matching on $B'$ and $A'$ and $|M''|=|M'|+1$. From the induction hypothesis and equation after first iteration we see that $|M|=|M''| = |M'|+1 \le |( produce\_UM \, B'\, A' )| + 1 = |(produce\_UM \, B\, A)|$. Observe that when $m_1=m_2$ we have $M''=(b,a,p)::(M \setminus \{m_1\})$ and $M'=(M \setminus \{m_1\})$.

\item Either $b$ or $a$ appears in bids or asks of $M$: Let $b$ appears in the bids of $M$ and $b=bid\_of \, m$ for some $m\in M$. Consider the matching $M''$ such that $M''=(b,a,p)::(M \setminus \{m\})$. The matching $M'=(M \setminus \{m\})$ is a uniform matching on $B'$ and $A'$ and $|M|=|M''| = |M'|+1 \le (| produce\_UM \, B'\, A' )| + 1 = |(produce\_UM \, B\, A)|$. The proof for the instance when $a$ appears in the asks of $M$ is identical to this.

\item Neither $b$ nor $a$ appears in bids or asks of $M$: Consider the matching $M''$ such that $M''=(b,a,p)::M$. The matching $M$ is a uniform matching on $B'$ and $A'$ and $|M| < |M''| = |M|+1 \le |( produce\_UM \, B'\, A' )| + 1 = |(produce\_UM \, B\, A)|$.

\end{enumerate} 
\end{enumerate}

\end{document}